\documentclass[12pt]{article}
\usepackage{amssymb,amsmath}
\usepackage{cite}
\usepackage{verbatim}
\usepackage{graphicx}

\let\OLDthebibliography\thebibliography
\renewcommand\thebibliography[1]{
  \OLDthebibliography{#1}
  \setlength{\parskip}{0pt}
  \setlength{\itemsep}{0pt plus 0.3ex}
}

\newcommand{\dd}{\mbox{\rm d}}
\newcommand{\wg}{\wedge}
\newcommand{\gam}{\gamma}
\newcommand{\Gam}{\Gamma}
\newcommand{\sg}{\sigma}

\newcommand{\dg}{\dagger}
\newcommand{\ddg}{\ddagger}
\newcommand{\tl}{\tilde}
\newcommand{\ul}{\underline}

\newcommand{\DD}{\mbox{\rm D}}

\newcommand{\nnn}{\noindent}

\newcommand{\p}{\partial}

\newcommand{\be}{\begin{equation}}
\newcommand{\bear}{\begin{eqnarray}}
\newcommand{\ear}{\end{eqnarray}}
\newcommand{\ee}{\end{equation}}
\newcommand{\lbl}{\label}
\newcommand{\bi}{\bibitem}
\newcommand{\ci}{\cite}

\newcommand{\vs}{\vspace}
\newcommand{\hs}{\hspace}

\begin{document}

\

\baselineskip .7cm 

\vs{8mm}

\begin{center}

{\LARGE Clifford Algebras, Spinors and $Cl(8,8)$ Unification}

\vs{3mm}

Matej Pav\v si\v c

Jo\v zef Stefan Institute, Jamova 39,
1000 Ljubljana, Slovenia

e-mail: matej.pavsic@ijs.si

\vs{6mm}

{\bf Abstract}
\end{center}

\baselineskip .5cm 

{\footnotesize 
	It is shown how the vector space $V_{8,8}$ arises from the Clifford algebra $Cl(1,3)$ of spacetime.
	The latter algebra describes fundamental objects such as strings and branes in terms of their
	$r$-volume degrees of freedom, $x^{\mu_1 \mu_2 ...\mu_r}$ $\equiv x^M$, $r=0,1,2,3$, that generalize
	the concept of center of mass. Taking into account that there are sixteen $x^M$, $M=1,2,3,...,16$, and in
	general $16 \times 15/2 = 120$ rotations of the form $x'^M = {R^M}_N x^N$, we can consider $x^M$
	as components of a vector $X=x^M q_M$, where $q_M$ are generators of the Clifford algebra $Cl(8,8)$.
	The vector space $V_{8,8}$ has enough room for the unification of the fundamental particles and forces of the
	standard model. The rotations in $V_{8,8}\otimes \mathbb{C}$ contain the grand unification group $SO(10)$ as a subgroup, and also the Lorentz group $SO(1,3)$. It is shown how the Coleman-Mandula no go theorem can be avoided.
	Spinors in $V_{8,8}\otimes \mathbb{C}$ are constructed in terms of the wedge products of the basis vectors rewritten in the Witt basis. They satisfy the massless Dirac equation in $M_{8,8}$ with the internal part of the Dirac operator giving the non vanishing masses in four dimensions.
}

\baselineskip .6cm

\section{Introduction}

Unification of fundamental interactions is still an unfinished project in theoretical physics.
Among many attempts string theory has for long time been a very promising
avenue. Strings are extended objects that upon quantization ``miraculously'' give
rise to gauge fields. Namely, the quantum excitations of a string contain Yang-Mills fields
or gravity, depending on whether the string is open or closed. In other words,
the interaction fields are associated with a string's (quantum) configurations.

Another possible approach to the unification employs the idea that the arena for physics is not
spacetime, but the configuration space of matter\ci{PavsicBook1,PavsicBled,PavsicIARD2016,PavsicBook2}. A matter configuration can be modeled
as a multiparticle system represented as a point in configuration space ${\cal C}$, whose
dimension equals the number of particles multiplied by the dimension of spacetime $M_{1,3}$.
In such scenario, spacetime is a subspace of configuration space,
associated with a chosen single particle. One can then proceed \`a la Kaluza-Klein
and obtain 4D gravity and gauge fields from the higher dimensional gravity in ${\cal C}$\ci{PavsicBled}.

This, among others, also brings an insight why a string configuration contains gauge
interaction fields. In string theory gauge fields occur at the quantum level, but according to the
postulated general relativity\ci{PavsicBook1,PavsicBled,PavsicIARD2016,PavsicBook2} in configuration space, they occur already
at the classical level for any matter configuration and therefore also for a string.

A string as an extended  object has infinite dimensions and can be represented as a point
in an infinite dimensional configuration space ${\cal C}$. Choosing a point $P$ on the string,
e.g., its center of mass, or one of its ends, it should then in principle be possible
to perform a Kaluza-Klein reduction from ${\cal C}$ to $M_{1,3}$, association with $P$,
and so obtain 4D gravity and gauge fields.

In the latter description we assumed that the string is embedded in 4-dimensional
spacetime. This is possible at the classical level, while in the quantized theory one
encounters inconsistencies which disappear in 26 dimensions. A peculiar feature
of a string is that though extended, it is still singular, because it is infinitely thin.
This indicates that a string is an idealized description of an actual physical object.
Exact strings, as well as exact point particles, do not exist in nature. Physical objects
are not exactly point-like nor exactly string-like. They have thickness. In
Refs.\ci{CastroChaos,CastroAurilia,PavsicBook1,Aurilia,PavsicArena,CastroPavsicReview,PavsicKaluza,PavsicKaluzaLong,PavsicLicata,PavsicMaxwellBrane}
it was explained how a physical object can be described not only in terms of its center of mass
coordinates, i.e., the coordinates of a point, but can also be sampled in more detail by
means of oriented lengths, areas, volumes and 4-volumes, in general, $r$-volumes, $r=0,1,2,3,4$, also called $r$-areas.
 Convenient mathematical objects that describe $r$-volumes are
Clifford numbers, the elements of a Clifford algebra, in our case the Clifford algebra
$Cl(1,3)$ of spacetime.
An extended objects can thus be described by $2^4=16$ degrees of freedom. By starting
with spacetime and considering the Clifford algebra over it we thus gain a lot of new
room for description of particles and the interactions between them. Namely, a $Cl(1,3)$
can be considered as a tangent space at a given point of a 16-dimensional manifold,
called {\it Clifford space} $C$ which, in general, can be curved. Employing the Kaluza-Klein
recipe, we thus obtain the unification of interactions in Clifford space\ci{PavsicKaluza,PavsicKaluzaLong}.

Once we have a Clifford space $C$, we can consider it as a manifold of ``its own''\ci{PavsicE8}
and forget that
we have arrived at it via Clifford algebra $Cl(1,3)$. Thus we just have a 16-dimensional
manifold $M_{8,8}$, such that the tangent space at any of its points is a vector space $V_{8,8}$.
By considering the rotations in $V_{8,8}$ we thus have the orthogonal group $SO(8,8)$
whose subgroups are $SO(6,4)$ and $SO(2,4)$.
The group Spin(6,4), the cover group of $SO(6,4)$,  is isomorphic to the Pati-Salam\ci{Pati-Salam,BaezPatiSalam} grand unification group $SU(4)\times SU(4)$.  If we consider complex valued spinors generated by basis vectors
of the vector space $V_{6,4}$, then we can equivalently consider complex valued spinors generated by, e.g.,
the vectors of $V_{0,10}$ that we denote $V_{10}$. The group $SO(10)$ of rotations in $V_{10}$
is considered, besides $SU(4)\times SU(4)$ and $SU(5)$, in grand unification of particles and forces (see the instructive review by Baez\ci{BaezPatiSalam}). 
Concerning the group $SO(2,4)$, it is related to the
Stueckelberg theory\ci{Stueckelberg1,Stueckelberg2,Horwitz1,Fanchi} with evolution parameter and to the two times physics considered by Bars\ci{Bars}. It contains the Lorentz group $SO(1,3)$ as a subgroup.
In this paper we consider the spinors of only one minimal left ideal of $Cl(8,8)$ and show that such a setup
incorporates grand unified theories. Consideration of the full $Cl(8,8)$ or its subgroup $Cl(8)$ brings a lot of additional possibilities concerning the unification\ci{Lisi1,Lisi2,Gillard,Gresnigt,CastroUnif1,CastroUnif2,CastroUnif3,CastroUnif4}.

\section{From the distances in spacetime to Clifford space: How a 16D space is ``embedded''
in a 4D space}

Spacetime consists of point events. The squared distance between two infinitesimally close
events is
\be
  \dd s^2 = \eta_{\mu \nu} \dd x^\mu \dd x^\nu~, ~~~~~\mu, \nu = 0,1,2,3 ,
\lbl{2.1}
\ee
There are two possible ways of taking the square root of the quadratic form (\ref{2.1}):
\be
 (i) ~~~\sqrt{\dd s^2} = \dd s = \sqrt{\eta_{\mu \nu} \dd x^\mu \dd x^\nu} ,
\lbl{2.2}
\ee
where $\dd s$ is the infinitesimal {\it scalar distance};
\be
  (ii) ~~~\sqrt{\dd s^2} = \dd x = \dd x^\mu \gam_\mu , \hs{1cm}
\lbl{2.3}
\ee
Here $\dd x$ is the infinitesimal {\it vector} that joints the point with the coordinates $x^\mu$ and
$x^\mu + \dd x^\mu$, while $\gam_\mu$ are basis vectors satisfying
\be
  \gam_\mu \cdot \gam_\mu \equiv \frac{1}{2} (\gam_\mu \gam_\nu + \gam_\nu \gam_\mu ) = \eta_{\mu \nu}.
\lbl{2.4}
\ee
The latter relation defines the generators $\gam_\mu$ of the Clifford algebra Cl(1,3) of spacetime $M_{1,3}$.

The vector (\ref{2.3}) is an oriented line element.  By wedge products of vectors we can form oriented areas, volumes and
4-volumes:
\be
  \dd x \wg \dd x' = \frac{1}{2} \dd x^{\mu \nu} \gam_\mu \wg \gam_\nu,
\lbl{2.5a}
\ee
\be
  \dd x \wg \dd x' \wg \dd x''= \frac{1}{3!} \dd x^{\mu \nu \rho } \gam_\mu \wg \gam_\nu \wg \gam_\rho ,
\lbl{2.5b}
\ee
\be
  \dd x \wg \dd x' \wg \dd x'' \wg \dd x'' 
   = \frac{1}{4!} \dd x^{\mu \nu \rho \sg} \gam_\mu \wg \gam_\nu \wg \gam_\rho \wg \gam_\sg ,
\lbl{2.5c}
\ee
where $\dd x^{\mu \nu}$, $\dd x^{\mu \nu \rho}$ and $\dd x^{\mu \nu \rho \sg}$ are the antisymmetrized
products of the vector components $\dd x^\mu$, $\dd x'^\mu$, $\dd x''^\mu$, $\dd x'''^\mu$, namely,
\be
 \dd x^{\mu \nu} = \dd x^\mu \dd x'^\nu - \dd x'^\mu \dd x^\nu,
\lbl{2.5d}
\ee
\be
 \dd x^{\mu \nu \rho} = \dd x^\mu \dd x'^\nu \dd x''^\rho + \dd x'^\mu \dd x''^\nu \dd x^\rho + \dd x''^\mu \dd x^\nu \dd x'^\rho
   - \dd x^\mu \dd x''^\nu \dd x''^\rho -\dd x''^\mu \dd x'^\nu \dd x^\rho -\dd x'^\mu \dd x^\nu \dd x''^\rho ,
\lbl{2.5e}
\ee
and the analogous expression for $\dd x^{\mu \nu \rho \sg}$.

The wedge product denotes the antisymmetrized product of vectors. For the basis vectors we have:
\be
   \gam_\mu \wg \gam_\nu \equiv \frac{1}{2} \left (\gam_\mu \gam_\nu - \gam_\nu \gam_\mu \right ),
\lbl{2.6}
\ee
\be
 \gam_\mu \wg \gam_\nu \wg \gam_\rho \equiv \frac{1}{3!} \left (\gam_\mu \gam_\nu \gam_\rho
 +\gam_\nu \gam_\rho \gam_\mu + \gam_\rho \gam_\mu \gam_\nu
    - \gam_\mu \gam_\rho \gam_\nu - \gam_\rho \gam_\nu \gam_\mu - \gam_\nu \gam_\mu \gam_\rho \right ),
\lbl{2.7}
\ee
and analogous for $\gam_\mu \wg \gam_\nu \wg \gam_\rho \wg \gam_\sg$.

In a manifold there are infinitely many ways of connecting its point into submanifolds, for instance, into surfaces of a lesser dimensionality. In particular, the submanifolds can be
associated with physical objects in spacetime. For instance,  a set of points can lie on a closed string.  Another possibility
is the set of points lying on an open membrane whose boundary is a closed loop. An infinitesimal oriented area
of that membrane (Fig.\,1) is the wedge product of two tangent vectors $ \dd \Sigma = \dd \xi_1 \wg \dd \xi_2$.
Expanding $\dd \xi_1$ and $\dd \xi_2$ in terms of the tangent basis vectors $e_a$, $a=1,2$, on the membrane, we have
\be
  \dd \Sigma = \dd \xi_1 \wg \dd \xi_2 = \dd \xi_1^a \dd \xi_2^b e_a \wg e_b = \frac{1}{2} \dd \xi^{ab} e_a \wg e_b ,
\lbl{2.7a}
\ee
where $\dd \xi^{ab} = \dd \xi_1^a \dd \xi_2^b - \dd \xi_2^a \dd \xi_1^b$. Recalling now that using the membrane's embedding
functions $X^\mu (\xi^a)$, we have
\be
   \dd x^\mu = \p_a X^\mu \dd \xi^a,
\lbl{2.7b}
\ee
where $\p_a \equiv \p/\p \xi^a$. Multiplying the right and the left side of Eq.\,(\ref{2.7b}) with $\gam_\mu$,
we obtain
\be
  \dd x = \dd x^\mu \gam_\mu = \p_a X^\mu \dd \xi^a \gam_\mu = \dd \xi^a e_a = \dd \xi,
\lbl{2.7c}
\ee
from which we have the the following relation between the spacetime basis vectors $\gam_\mu$ and the brane's basis
vectors $e_a$:
\be
  e_a = \p_a X^\mu \gam_\mu.
\lbl{2.7d}
\ee
Inserting (\ref{2.7d}) into Eq.\,(\ref{2.7a}) and integrating the infinitesimal oriented surface elements
$\dd \Sigma$ over the membrane, we obtain\ci{PavsicArena}
\be
  \int_\Sigma \dd \Sigma = \frac{1}{4} \int_\Sigma \dd \xi^{ab}
   \left (\p_a X^\mu \p_b X^\nu - \p_a X^\nu \p_b X^\mu \right ) \gam_\mu \wg \gam_\nu
   = \frac{1}{2} x^{\mu \nu}  \gam_\mu \wg \gam_\nu ,
\lbl{2.8}
\ee
where
\be
  x^{\mu \nu} =  \frac{1}{2} \int_\Sigma \dd \xi^{ab}
  \left (\p_a X^\mu \p_b X^\nu - \p_a X^\nu \p_b X^\mu \right ) .
\lbl{2.9}
\ee
By the Stokes theorem, Eq.\,(\ref{2.9}) becomes
\be
    x^{\mu \nu} =  \frac{1}{2} \int_B \dd s
  \left (X^\mu \frac{\p X^\nu}{\p s} - X^\nu \frac{\p X^\mu}{\p s} \right ) ,
\lbl{2.10}
\ee 
where $X^\mu (s)$ are the embedding functions of the boundary loop $B$, $s$ being a
parameter along the loop.
\setlength{\unitlength}{.68mm}

The quantity $x^{\mu \nu}$ denotes the effective oriented area of the surface $\Sigma$
bounded by the loop $B$. From the relation (\ref{2.10}) we see that $X^{\mu \nu}$
depends only on the boundary $B$ and is independent on the shape of $\Sigma$.
Eqs.\,(\ref{2.9}) and (\ref{2.10}) tell us that we have a mapping
\be
  X^\mu (\xi^a) \longrightarrow x^{\mu \nu},
\lbl{2.11}
\ee
or equivalently,
\be
   X^\mu (s) \longrightarrow x^{\mu \nu},
\lbl{2.12}
\ee
from an infinite dimensional object, namely a surface $\Sigma$, described by $X^\mu (\xi^a)$,
or its boundary $B$, described by $X^\mu (s)$, into the finite dimensional object
$x^{\mu \nu}$.

\setlength{\unitlength}{.68mm}
\begin{figure}[h!]
	
	\centerline{\includegraphics[scale=0.55]{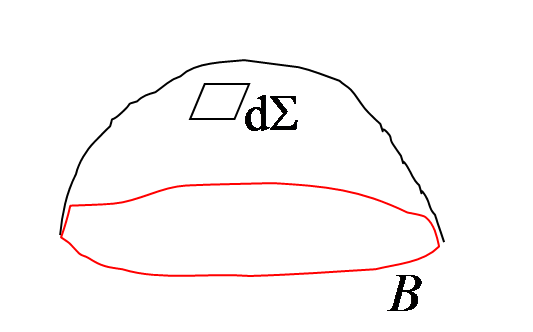}}
	
	\caption{\footnotesize An aread element $\dd \Sigma$ on a surface with the boundary $B$.}
\end{figure}

Such arrangement can describe two physically distinct objects\ci{PavsicLocalTachyons}:
\begin{description}
 \item	\ (i) A loop $B$ can be a closed instantonic string. Then $x^{\mu \nu}$ are the bivector coordinates associated with the closed instantonic string.
 \item (ii) A surface $\Sigma$ can correspond to an open instantonic 2-brane
 whose boundary is $B$.
 Then $x^{\mu \nu}$ are bivector coordinates associated with the open instantonic 2-brane.
\end{description}

Analogous holds for objects of arbitrary dimension. The corresponding multivector
 ($r$-vector) coordinates are then
$$
  x^{\mu_1\mu_2 ...\mu_r} = \frac{1}{r!} \int \dd \xi^{a_1 a_2 ...a_r}
  \p_{a_1} X^{[\mu_1} \p_{a_1} X^{\mu_2} ...\p_{a_r} X^{\mu_r ]}  \hs{2cm}$$
\be
 \hs{2cm} =  \frac{1}{r!} \int \dd s^{\ul a_1 \ul a_2 ...\ul a_{r-1}}
       X^{[\mu_1} \p_{\ul a_1} X^{\mu_2} ...\p_{\ul a_{r-1}} X^{\mu_r ]} 
\lbl{2.13}
\ee
Here
$\dd \xi^{a_1 a_2 ...a_r} \equiv \dd \xi^{[a_1} \dd \xi^{a_2}...\dd \xi^{a_r ]}$
and $\dd s^{\ul a_1 \ul a_2 ...\ul a_r}
 \equiv \dd s^{[\ul a_1} \dd s^{\ul a_2}...\dd s^{\ul a_r]}$
are, respectively, the infinitesimal elements of an $r$-dimensional surface and of its $(r-1)$-dimensional
boundary. The parameters (coordinates) denoting the points in those manifolds are $\xi^a$, $a=1,2,...,r$, and $s^{\ul a}$, $\ul a =1,2,...,r-1$, while $\p_a \equiv \p/\p \xi^a$ and $\p_{\ul a} \equiv
\p/\p s^{\ul a}$ are the derivatives with respect to those parameters.
The bracket `$[\ ]$' denotes antisymmetrization of the expression.

The functions $X^\mu (\xi^a)$, $a=1,2,...,r$, can describe an $r$-dimensional surface,
shortly $r$-surface, bounded by an $(r-1)$-dimensional surface. Equation (\ref{2.13})
determines the mapping from an infinite dimensional object, namely an $r$-surface,
$X^\mu (\xi^a)$, $a=1,2,...,r$, or its boundary, $X^\mu(s^{\ul a})$, $\ul a = 1,2,...,r-1$,
into the finite dimensional object $x^{\mu_1,\mu_2,...,\mu_r}$.
	
The quantity $x^{\mu_1,\mu_2,...,\mu_r}$ can thus describe\ci{PavsicBg2008,PavsicLocalTachyons}
 two distinct types of physical objects in spacetime\footnote{
	The objects considered here are point ``particles'' (events), strings, membranes,
	in general, branes,
	in {\it spacetime}. Usually those names refer to the objects in {\it space} (that is, in a three dimensional subspace of spacetime) that in spacetime sweep a worlsline, a worldsheet, a worlvolume, etc.. An $r$-brane considered here does not sweep an $(r+1)$-dimensional worldsheet, therefore we call it {\it instantonic} $r$-brane. Once this nomenclature is clear, we can omit ``instantonic''.}
    
a) an open instantonic $r$-brane;

b) a closed instantonic $(r-1)$-brane.

\nnn The mapping is many--to--one, so that $x^{\mu_1,\mu_2,...,\mu_r}$ is associated
with a class of those physical objects, whose representative is an {\it oppen  instantonic}
$r$-brane or, alternatively, a {\it closed instantonic} $(r-1)$-brane.

Distinction between those two types of objects can be formally made by means of a scalar
parameter $\sg$ which in the case of an open string is proportional to its length,
in the case of an open membrane (2-brane) to its scalar area, and, in general, the
scalar $r$-volume ($r$-area) of an open $r$-brane. For a closed $(r-1)$-brane the scalar
parameter $\sg$ vanishes. For instance, in the case of a closed string there is no
``material'' embraced by the string, therefore the scalar area $\sg$ associated with such
string is zero\footnote{
	Recall that the corresponding 2-vector (oriented) area of a closed string is different from zero and given by the bivector $X^{\mu \nu} \gam \wg \gam_\nu$.}.
This is not so so for an open membrane and, in general, for an open $r$-brane. Then
\be
  \sg = \frac{1}{A_r} \int \dd \xi^1 \dd \xi^2 ... \dd \xi^r \left ( {\rm det}
   \frac{\p X^\mu}{\p \xi^a} \frac{\p X^\nu}{\p \xi^b} \eta_{\mu \nu} \right )^{1/2},
   ~~~a,b,1,2,...,r,
\lbl{2.15}
\ee
is different from zero. The quantity $A_r$, $r=1,2,3,4$,
is defined as\ci{PavsicLocalTachyons}
\be
  A_r = \sqrt{X^\ddg *X}~,~~~~~   X= x^{\mu_1 \mu_2...\mu_r} \gam_{\mu_1} \wg \gam_{\mu_2}...
        \wg \gam_{\mu_r} ,
\lbl{2.15a}
\ee
where $\sqrt{~~~}$ is the scalar square root of the expression. Here we have generalized the
case considered in Eqs.\,(\ref{2.1}) and (\ref{2.2}). The symbol $\ddg$ denotes reverions, i.e., the operation that reverses the order of vectors in an expression, e.g., $(\gam_\mu \gam_\nu)^\ddg = \gam_\nu \gam_\mu$; the star $*$ denotes the scalar product of two multivectors, defined as $A*B = \langle A B \rangle_0$, where $\langle ~~~ \rangle_0$ denotes
the scalar part of the expression.

For an {\it  open string} the multivector grade is $r=1$ and the formula (\ref{2.13}) gives
\be
  X^\mu = X_2^\mu - X_1^\mu ,
\lbl{2.16}
\ee
where $X_1^\mu$ and $X_2^\mu$ are the coordinates of the string's ends. Equation (\ref{2.16})
is the result of the integration of the oriented line elements (\ref{2.3}), i.e, the infinitesimal vectors, along the string.The result of such integration is a finite vector
$X^\mu \gam_\mu$, whose components are given in Eq.\,(\ref{2.16}).

The scalar associated with an open string is determined by Eq.\,(\ref{2.15}) for $r=1$:
\be
  \sg= \frac{1}{A_1} \int \dd \xi \left (\frac{\p X^\mu}{\p \xi}\frac{\p X^\nu}{\p \xi}
  \eta_{\mu \nu} \right )^{1/2},
\lbl{2.17}
\ee
where
\be
  A_1 = \sqrt{(X^\mu \gam_\mu)^2} = \sqrt{X^\mu X^\nu \eta_{\mu \nu}} \equiv
  	\sqrt{( X_2^\mu - X_1^\mu)( X_2^\nu - X_1^\nu) \eta_{\mu \nu}} ,
\lbl{2.17a}
\ee
which, in general, is differrent from the string length
$\int \dd \xi \left (\frac{\p X^\mu}{\p \xi}\frac{\p X^\nu}{\p \xi}
\eta_{\mu \nu} \right )^{1/2}$.
This scalar $\sg$ is obtained if we integrate the infinitesimal scalar distances (line
elements) along the string.

Let us now take into account that a string is an idealization and that the
actual object is not a string but a thick string.
Then also the 2-vector, 3-vector and 4-vector coordinates $X^{\mu \nu}$, $X^{\mu \nu \rho}$,
can be different from zero.

The multivector coordinates $X^{\mu_1,\mu_2,...,\mu_r}$, $r=0,1,2,3,4$, describing an
extended object in spacetime, are components of a Clifford number
$$
  X = \sg \ul 1 + x^\mu \gam _\mu + \frac{1}{2} x^{\mu \nu} \gam_\mu \wg \gam_\nu
  + \frac{1}{3!} x^{\mu \nu \rho } \gam_\mu \wg \gam_\nu \wg \gam_\rho 
  + \frac{1}{4!} x^{\mu \nu \rho \sg } \gam_\mu \wg \gam_\nu \wg \gam_\rho \wg \gam_\sg	$$
\be
 = \sum_{r=0}^4 x^{\mu_1 \mu_2 ... \mu_r} \gam_{\mu_1 \mu_2 ... \mu_r}
  \equiv \sum x^M \gam_M \equiv x^M \gam_M  ~~(\text{Einstein's summation convention}).
\lbl{2.18}
\ee
Here $\gam_{\mu_1 \mu_2 ... \mu_r} \equiv \gam_{\mu_1} \wg \gam_{\mu_2}
\wg ... \wg \gam_{\mu_r}|_{\mu_1 < \mu_2 < ... < \mu_r}$,
therefore the factor
$1/r!$ does not take place in the expansion of $X$ in terms of $\gam_{\mu_1 \mu_2 ... \mu_r}$.
In the last step we have condensed the notation even more by introducing $\gam_M 
\equiv \gam_{\mu_1 \mu_2 ... \mu_r}$ and $x^M
 \equiv x^{\mu_1 \mu_2 ... \mu_r}|_{\mu_1 < \mu_2 < ... < \mu_r}$.

The Clifford number $X$, the so called {\it polyvector}, is the sum of multivectors of the
grades from $r=0$ to $r=4$. It denotes position in the 16-dimensional {Clifford space} $C$,
a manifold that has at any of its points the Clifford algebra $Cl(4)$, more precisely,
$Cl(1,3)$, as the tangent space.

The objects described by $x^{\mu_1 \mu_2 ... \mu_r}$ are ``instantonic'' in the sense that they occupy 
a {\it finite} region of spacetime, so that they are localized not only in 3-space but
also in time. They are not infinitely extended along a time-like direction as in the case
of a particle's worldline or a string's worldsheet.

When we talk about a ``particle'' we usually have in mind an object that looks like
a point in 3-space, and as a line in 4D spacetime $M_4$. Similarly, by ``string''
we usually mean anobject whose spatial form is a string, while it is a 2-dimensional
sheet in $M_4$.

In this work we consider the concept of {\it instantonic string}, {\it instantonic
membrane}, and, in general, an {\it instantonic $r$-brane}, $r=0,1,2,3,4$ which
generalize the concept of {\it event} in spacetime. The configuration
spaces of those objects, described by the embedding functions $X^\mu (\xi^a)$,
$a=1,2,3,4$, are infinite dimensional. The description of $r$-branes in terms
of the multivector coordinates $X^{\mu_1,\mu_2,...,\mu_r}$, $r=0,1,2,3,4$, is achieved
by the mapping
\be
  X^\mu (\xi^a) \longrightarrow x^{\mu_1 \mu_2 ...\mu_r}~,~~~~~~\mu=0,1,2,3;~~a=1,2,...,n; ~~~~n=1,2,3,4,
\lbl{3.19}
\ee
according to Eqs.\,(\ref{2.13}), (\ref{2.15}). The cases $n=1,2,3,4$ denote, respectively,
an instantonic string, instantonic membrane, instantonic 3-brane and instantonic 4-brane.
The latter object fils a 4-dimensional region of spacetime; it is like a spacetime
filling brane that not fills all but only a portion of spacetime.

The description of an instantonic object by a polyvector $X$, as defined in Eq.\,(\ref{2.18}),
comprises the usual $p$-branes, $p=1,2,3$, as limiting cases in which their time-like
extension goes to infinity.

A polyvector $X=x^M \gam_M$, $M = 1,2,...,16$, thus encodes a configuration of an
extended object, not in all its infinite detail, but in terms of the oriented
$r$-areas associated with the object

\section{Clifford space as the arena for physics}

The dynamics of the objects described by polyvectors has been investigated in
Refs.\ci{PavsicMaxwellBrane,PavsicBg2008}, where the concept of relativity in Clifford space ($C$-space)
was developed\ci{CastroChaos,PavsicBook1,CastroPavsicReview,PavsicKaluzaLong,
PavsicMaxwellBrane} and pointed out how it leads to the unification of particles and forces\ci{PavsicKaluza,PavsicKaluzaLong,CastroUnific}.
This is possible because the arena for physics is taken to be the 16-dimensional
Clifford space $C$. The points of $C$ correspond to the instantonic extended objects
in 4D spacetime modeled by the polyvector coordinates $x^{\mu_1 \mu_2 ... \mu_r} \equiv x^M$.

Taking the differential of the polyvector (\ref{2.18}),
\be
  \dd X = \dd x^M \gam_M ,
\lbl{3.1a}
\ee
the quadratic form in Clifford space is,
\be
  \dd S^2 = \dd X^\ddg * \dd X = \dd x^M \dd x^N \eta_{MN},
\lbl{3.1}
\ee
where the metric is
\be
  \eta_{MN} = \gam_M^\ddg * \gam_N = \langle \gam_M^\ddg \gam_N \rangle_0 .
\lbl{3.2}
\ee
With such a definition of the metric, the signature is $(8,8)$, so that the
explicit form of the expression (\ref{3.1}) is
\bear
  &&\dd S^2 = \dd \sg^2 + (\dd x^0)^2 - (\dd x^1)^2 - (\dd x^2)^2 - (\dd x^3)^2 \nonumber \\
 && \hs{1cm} -(\dd x^{01})^2 - (\dd x^{02})^2 - (\dd x^{03})^2 + (\dd x^{12})^2 + (\dd x^{13})^2 +(\dd x^{23})^2 \nonumber \\
 &&\hs{1cm}- (\dd {\tl x}^0)^2 + (\dd {\tl x}^1)^2 + (\dd {\tl x}^2)^2+(\dd {\tl x}^3)^2 - \dd {\tl \sg}^2 ,
\lbl{3.3}
\ear
which comes after rewriting the polyvector (\ref{2.18}) according to
\be
  X = \sg {\ul 1} + x^\mu \gam_\mu + \frac{1}{2} x^{\mu \nu} + {\tl x}^\mu \gam_5 \gam_\mu
  + {\tl \sg} \gam_5 ,
\lbl{3.4}
\ee
where
\be
  {\tl x}^\mu = \frac{1}{3!} {\epsilon^\mu}_{\nu \rho \sg} x^{\nu \rho \sg}~,
  ~~~~~~~{\tl \sg} = \frac{1}{4!} \epsilon_{\mu \nu \rho \sg} x^{\mu \nu \rho \sg}~,
\lbl{3.5}
\ee
\be
  \gam_\mu \wg \gam_\nu \wg \gam_\rho = {\epsilon_{\mu \nu \rho}}^\sg \gam_5 \gam_\sg ~,
 ~~~~~~~~ \gam_\mu \wg \gam_\nu \wg \gam_\rho \wg \gam_\sg = \epsilon_{\mu \nu \rho\sg} \gam_5 .
\lbl{3.6}
\ee

A worldline in $C$, described by the equation $x^M = X^M (\tau)$,  represents the
evolution of a  `thick’ particle in spacetime (Fig.\,2). Thick particle can be an
aggregate of $p$-branes for various $p=0,1,2,…$. But such interpretation is not obligatory. Thick particle may be a conglomerate of whatever extended objects that can be sampled
by polyvector coordinates $X^M  \equiv X\,^{\mu _1 \mu _2 ...\mu _r }$, $r=0,1,2,3$.
The action for such a system is
\be
I = \kappa \int_{}^{} {\dd \tau \,(\eta _{MN} \dot X^M \dot X^N } )^{1/2}, 
\lbl{3.7}
\ee
where $\kappa$ is a constant having the role of mass in $C$.
This is just like the point particle action, only that the ``point particle'' is now
in 16-dimensional Clifford space $C$. The equation of motion
\be
\ddot X^M \, \equiv \,\,\frac{{\,{\rm{d}}^{\rm{2}} X^M }}{{{\rm{d}}\tau ^2 }}\,\, = \,\,0
\lbl{3.8}
\ee
describes a flat worldline in $C$, which corresponds to a tensionless brane in $M_4$.
For the branes with tension one has to introduce
curved Clifford space in which instead of the flat space metric $\eta_{MN}$
we take a curved metric $g_{MN}$. Then Eq.\,(\ref{3.8}) generalizes to the equation of
a geodesic in $C$. For a particular choice of metrc\ci{PavsicBg2008} one obtains the description of the Dirac-Nambu-Goto brane sampled the coordinates
$X^M  \equiv X\,^{\mu _1 \mu _2 ...\mu _r }$.

\setlength{\unitlength}{.68mm}
\begin{figure}[h]
	
	\centerline{\includegraphics[scale=0.45]{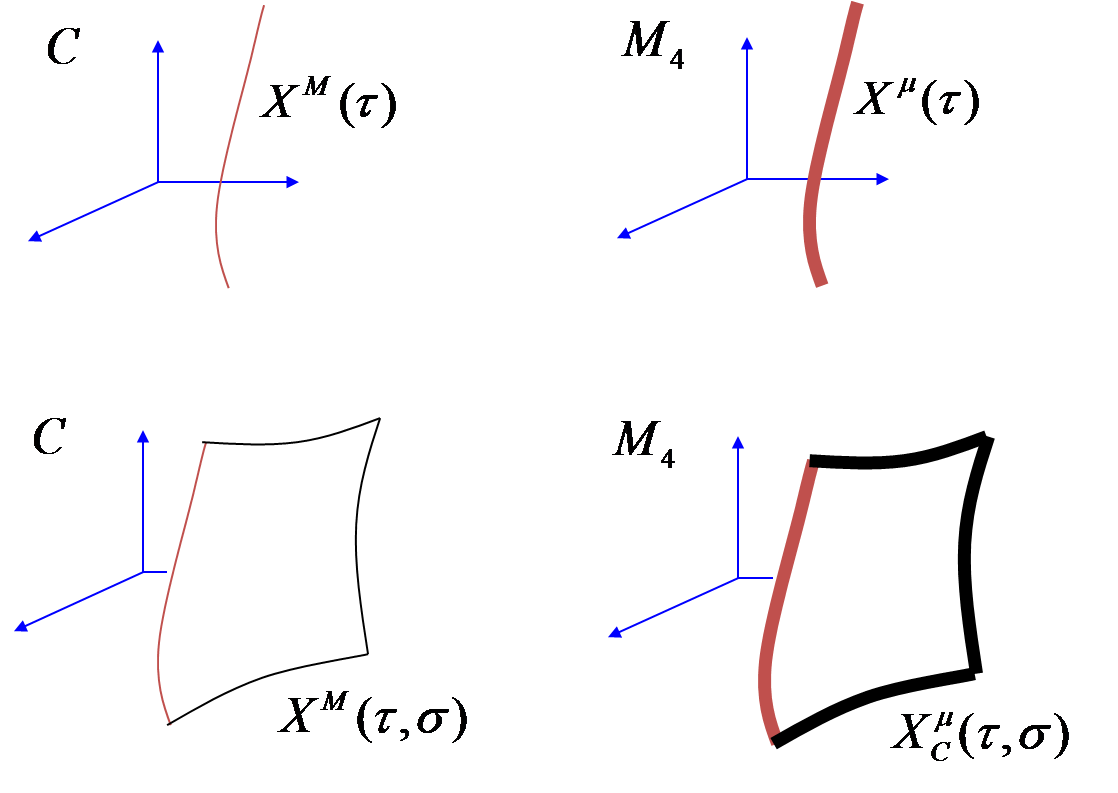}}
	
	\caption{\footnotesize A worldline in Clifford space $C$ corresponds to a thick worldline in
	spacetime $M_4$ (up). A wordlsheet in $C$ corresponds to a thick worldsheet in $M_4$ (down).}
\end{figure}

A world sheet in $C$, described by the equation $x^M = X^M (\tau,\sg)$, represents
the evolution of  a  `thick’ string in spacetime $M_4$ (Fig.\,2). Thick string can be
an aggregate $p$-branes for various $p=0,1,2,…$. But such interpretation is not obligatory.
Thick string may be a conglomerate of whatever extended objects that can be sampled
by polyvector coordinates $X^M  \equiv X\,^{\mu _1 \mu _2 ...\mu _r }$. The action
for a $C$-space string in conformal gauge is
\be
I = \frac{\kappa }{2}\int \dd \tau \,\dd \sigma \,({\dot X}^M {\dot X}^N  - X'^M  X'^N )\,\eta _{MN} 
\lbl{3.9}
\ee
The space in which such string lives is Clifford space. Its dimension is 16, and signature $(8,8)$.
In Ref.\ci{PavsicSaasFee} it was shown that upon quantization there are no central terms in the Virasoro algebra, if the space in which the
string lives has signature $(8,8)$, provided that the definition of vacuum as considered in Refs.\ci{Kim,Cangemi,KimBook,Jackiw,PavsicPseudoHarm} is used.
According to Refs.\ci{PavsicPseudoHarm,Woodard}, such vacuum definition gives the correct quantization of the
harmonic oscillator in a pseudo-Euclidean space, because it gives the correct classical limit.

\section{Promoting the Clifford algebra $Cl(1,3)$ to a vector space $V_{8,8}$}

Clifford algebra is a vector space. In particular, $Cl(1,3)$ is a 16D vector space with signature $(8,8)$.
Let us denote it $V_{8,8}$. We will now adopt the view of Ref.\ci{PavsicE8}
and consider $V_{8,8}$ as a vector space
spanned by the basis vectors $q_M$ that satisfy
\be
  q_M \cdot\,q_N \, = \frac{1}{2}\left( {q_M q_N \, + q_N q_M } \right)\, = \,\eta _{MN} ~,~~~~~~~
  M,N=1,2,3,...,16,
\lbl{4.1}
\ee
where, as before,  $\eta_{MN} = {\rm diag} (1,1,1,1,1,1,1,1,-1,-1,-1,-1,-1,-1,-1,-1)$,
i.e., the metric with signature $(r,s) =(8,8)$.
Therefore, if, instead of the basis 
$\gam_M \equiv \gam_{\mu_1 \mu_2 ... \mu_r}$ of the Clifford algebra $Cl(1,3)$,
we take the generators $q_M$ of $Cl(8,8)$, satisfying Eq.\,(\ref{4.1}), we obtain the same
quadratic form (\ref{3.3}).

The $V_{8,8}$ is a tangent space to the 16-dimensional manifold $M_{8,8}$. In general one has to distinguish
between the coordinate frame field and the orthonormal frame field. Let us denote with $q_M$ the coordinate basis
vectors and with $q_A$, $A =1,2,3,...,16$, the orthonormal basis vectors. Here for simplicity we use the same 
symbol $q$ and distinguish those two different sorts of objects by the indices $M$ and $A$.

A possible local decomposition of $M_{8,8}$ is
\be
M_{8,8}  = M_{1,1} \,\dot  +\, M_{1,3}\, \dot  + \,M_{6,4} 
\lbl{4.2}
\ee

The subspace $ M_{1,3}$ is the Minkowski spacetime. The subspace $M_{1,1}$ together with $M_{1.3}$
forms the six-dimensional space $M_{2,4}$. Good features of $M_{2,4}$:

a) It is the arena for 2T physics (Bars\ci{Bars}).

b) It enables the Stueckelberg theory\ci{Stueckelberg1,Stueckelberg2,
	Horwitz1,Fanchi,PavsicBook1}.

c) It is the arena for the conformal group.

The subspace $M_{6,4}$ has the role of the internal space that brings into the game the additional interactions,
besides the gravity in 4D spacetime. 
A good feature of $M_{6,4}$ is that it serves as the arena for the SO(10) grand unification.

In this scheme the overall arena for physics is $M_{8,8}$. A point in $M_{8,8}$ has coordinates
$X^M$, $M = 1,2,...,16$.
 The group SO(8,8) acting within
a tangent space $V_{8,8}$ of $M_{8,8}$ contains:
\be
    SO(8,8) \supset \,SO(2,4) \times SO(6,4),
\lbl{4.3}
\ee
where
\be
    SO(2,4) \supset \,SO(1,3) \times \,SO(1,1)
\lbl{4.4}
\ee
Here $SO(1,3)$ is the Lorentz group which together with $SO(1,1)$ is a subset of the conformal
group $SO(2,4)$.
 
For the second factor in Eq.\,(\ref{4.4}) we have
\be
SO(6,4) \supset SO(6) \times SO(4) \leftarrow SU(4) \times SU(2) \times SU(2),
\lbl{4.5}
\ee
which gives the Pati-Salam unified model\ci{Pati-Salam,BaezPatiSalam}.
We have thus arrived at the framework which enables the unification of the Lorentz group with the
Pati-Salam group, a descent of which is the Standard model gauge group $SU(3) \times SU(2) \times U(1)$.

What about the Coleman-Mandula theorem\ci{Coleman-Mandula}? Its starting assumption is the unitarity of the
$S$-matrix in $M_{1,3}$,
which holds for a wave function that depends on position in spacetime. As a consequence, no mixing
of spacetime and internal symmetries is possible within such a setup. This is not true for a theory
whose starting point is a higher dimensional space, such as, e.g., the Clifford space or the space of
the Kaluza-Klein theory, or the space $M_{8,8}$ considered here, in which the wave function is a function
of all $N>4$ coordinates, and thus unitarity holds in  the higher dimensional space. Then, fundamentally, there is no distinction between external and internal
symmetries, and therefore they can mix among themselves. Only after a symmetry breaking, eg.,
in Kaluza-Klein theories due to isometries along the internal dimensions or a compactification of
small extra dimensions, a wave function is effectively a function of spacetime coordinates only.
Then, of course, for such effective theory in four dimensions one gets that the spacetime and 
internal symmetries do not mix. However, for an underlying more fundemental theory in a higher
dimensional space, the Coleman-Mandula theorem is not applicable. To sum up, the spacetime and the internal
symmetries can only not mix after a symmetry breaking which sets apart the 4D spacetime and
a higher dimensional "internal" space. That there is a loophole in the Coleman-Mandula theorem was shown
in Ref.\ci{Lisi2} by a different argumentation, namely, that before symmetry breaking,
there is no metric and thus no S-matrix.

{\it Vector fields in} $M_{8,8}$

Upon the (first) quantization of the classical system described by the action (\ref{3.7}) we obtain the Klein-Gordon equation in $M_{8,8}$:
\be
    (\partial ^M \partial _M - \kappa^2 )\Phi (X^M ) = 0~,~~~~M=1,2,3,...,16,
\lbl{4.6}
\ee
where $X^M$ denotes coordinates of position in $M_{8,8}$, and $\p_M \equiv \p/\p x^M$.
The latter equation comes from the classical momentum constraint $P_M P^M - \kappa^2=0$ which becomes the operator equation
acting on a state $\Phi$.

For a state we take a vector field in $M_{8,8}$, 
\be
   \Phi  = \phi ^A q_A~,~~~~~q_A \cdot q_B = \eta_{AB}~,~~~~~A,B=1,2,3,...,16.
\lbl{4.7}
\ee
If the components $\phi^A$ are complex valued, then we can choose a new basis vectors $q'^A$ such that the same
vector field $\Phi$ can be expanded in terms of those new basis vectors according to
   \be
   \Phi  = \phi'^A q'_A~,~~~~~q'_A \cdot q'_B = \eta'_{AB}~,~~~~~A,B=1,2,3,...,16,
   \lbl{4.8}
   \ee
where the signature of the new metric $\eta'_{AB}$ differs from the signature of the old metric $\eta_{AB}$.
The new signature, $(r,s)$, can be arbitrary. In particular it can be $r=2$, $s=14$, which corresponds to
\be
    V_{2,14} \, = \,V_{2,4}  {\dot +} V_{0,10} .
\lbl{4.9}
\ee
Instead of the basis of the original vector space $V_{8,8}$ we can thus take the basis of the space
$V_{2,14}$ that can be split into the space $V_{2,4}$ and $V_{0,10}$. Over the latter vector space
we can construct the spinors of the $SO(10)$ grand unification.

\vs{2mm}

{\it Spinor fields in} $M_{8,8}$

We have considered {\it complex vector fields} in the manifold $M_{8,8}$.  Let us now
consider {\it complex spinor fields} in $M_{8,8}$. At any of its points the tangent space is
$V_{8,8}$. Its basis vector $q_A$, $A=1,2,3,...,16$, can be split into the time-like and space-like part
according to 
\be
\begin{array}{l}
	q_A  = (q_a ,\tilde q_a )\,,\,\,\,\,\,\,\,\,\,\,a = 1,2,3,...,8\, \\ 
	\,\,\,\,\,\,\,\,\,\,\,\,\,q_a^\dag   = \,q_a \,,\,\,\,\,\,\,\,\,\,\,\,\,\,\,\,\,\,\,\,\,\,\,\,\,\,q_a \cdot q_b \, = \delta _{ab}  \\ 
	\,\,\,\,\,\,\,\,\,\,\,\,\,\tilde q_a^\dag   = \,\, - \tilde q_a \,,\,\,\,\,\,\,\,\,\,\,\,\,\,\,\,\,\,\,\,\,\,\tilde q_a \cdot\tilde q_b^{} \, = \, - \delta _{ab}  \\ 
\end{array}
\lbl{4.10}
\ee
	
An alternative basis is the Witt basis:
\be
\begin{array}{l}
\chi _a \, = \frac{1}{2}\,\left( {q_a  + \tilde q_a } \right) \\ 
\chi _a^\dag  \, = \frac{1}{2}\,\left( {q_a  - \tilde q_a } \right) \\ 
\end{array}
\lbl{4.11}
\ee
Writing $\tl q_a = i \bar q_a$ we have
\be
\begin{array}{l}
q_A  = (q_a ,i\,\bar q_a )\,,\,\,\,\,\,\,\,\,\,\,a = 1,2,3,...,8\, \\ 
\,\,\,\,\,\,\,\,\,\,\,\,\,q_a^\dag   = \,q_a \,,\,\,\,\,\,\,\,\,\,\,\,\,\,\,\,\,\,\,\,\,\,\,\,\,\,q_a \cdot q_b \, = \delta _{ab}  \\ 
\,\,\,\,\,\,\,\,\,\,\,\,\,\bar q_a^\dag   = \,\bar q_a \,,\,\,\,\,\,\,\,\,\,\,\,\,\,\,\,\,\,\,\,\,\,\,\,\,\bar q_a \cdot\bar q_b^{} \, = \,\delta _{ab}  
\end{array}
\lbl{4.12}
\ee
The Witt basis then reads
\be
\begin{array}{l}
\chi _a \, = \frac{1}{2}\,\left( {q_a  + \,i\,\bar q_a } \right) \\ 
\chi _a^\dag  \, = \frac{1}{2}\,\left( {q_a  - \,i\,\bar q_a } \right) 
\end{array}
\lbl{4.13}
\ee
It satisfies the fermionic anticommutation relations
\be
\{ \chi _a ,\chi _b^\dg \} \, = \,\delta _{ab} ~,~~~~\{ \chi _a ,\chi _b \} \, =\{ \chi _a^\dg ,\chi _b^\dag  \} \, =0.
\lbl{4.14}
\ee

Spinors are given in terms of the creation operators $\chi _a^\dg$ acting on
the vacuum\ci{SpinorFock,BudinichP,Giler,Winnberg,PavsicSpinorInverse,PavsicOrthoSymp,PavsicFirence,BudinichM}
\be
\Omega  = \prod\limits_{a = 1}^8 {\chi _a } 
\lbl{4.15}
\ee
A generic spinor field in $M_{8,8}$ is:
\be
\Psi  = \left( {\psi ^0 1 + \psi ^{a_1 } \chi _{a_1 }^\dag   + \psi ^{a_1 a_2 } \chi _{a_1 }^\dag  \chi _{a_2 }^\dag   + ...\psi ^{a_1 a_2 ...a_8 } \chi _{a_1 }^\dag  \chi _{a_2 }^\dag  ...\chi _{a_8 }^\dag  } \right)\Omega \equiv \psi^{\tl A} \xi_{\tl A},
\lbl{4.16}
\ee
which is a superposition of spinor components $\psi^{\tl A}$ and basis spinors $\xi_{\tl A}$.
We assume
that it depends on position in $M_{8,8}$ and satisfies the generalized Dirac equation	
\ci{PavsicKaluza,PavsicKaluzaLong,PavsicMoskva}:      
\be
\left ( i\gamma ^M \partial _M + \kappa \right )\Psi  = 0,
\lbl{4.17}
\ee
where $\kappa$ is the mass in sixteen dimensions. This is analogous to the usual procedure where spinors are assumed to satisfy the Dirac equation in $M_{1,3}$.

In general, the components $\psi ^{a_1 a_2 ...a_r }$, $r = 0,1,2,...,8$, are complex.
Therefore the same spinor  can be generated in terms of the basis vectors $q_A$ of any signature $p,q$.
For instance, instead of constructing spinors over
\be
V_{8,8}  \otimes \mathbb{C}
\lbl{4.18}
\ee
we can construct them over
\be
(V_{2,4} \, \dot + \,V_{0,10} ) \otimes \mathbb{C}.
\lbl{4.19}
\ee
In the subspace $V_{2,4}$ live the spinors of Minkowski space, whilst in $V_{0,10}$ live the spinors of $SO(10)$ grand unification.

Spinors in  $M_{8,8}$ are functions of position $X^M$ in $M_{8,8}$. They have values
(as members of minimal ideals) in the complexified $Cl(8,8)$ which can be written as
\be
\begin{array}{l}
Cl(8,8) \otimes C = Cl(2,14) \otimes C \\ 
\,\,\,\,\,\,\,\,\,\,\,\,\,\,\,\,\,\,\,\,\,\,\,\,\,\, = \left( {Cl(2,4) \otimes Cl(10)} \right) \otimes C\,\,\, = Cl(16) \otimes C \\ 
\end{array}
\lbl{4.20}
\ee

Crucial here are the two subalgebras:

  \  (i) In $Cl(2,4)$ are contained the spinors of Minkowski space, as a part of a larger theory
which can be, as mentioned before, the 2T physics by Bars, the Stueckelberg theory, or conformal theory.

   (ii) In $Cl(10)$ have values the spinors of $SO(10)$ grand unification.
   
In the following, instead of considering the full theory in $M_{8,8}$, we will confine us here
to the theory
in the subspace $M_{7,7}= M_{1,3} \dot + M_{6,4}$ and so neglect the piece $M_{1,1}$ which together
with spacetime forms $M_{2,4}$.

 A spinor field in $M_{7,7}$ can be writen as
\be
  \Psi  = \psi^{\tl A} \xi_{\tl A} \equiv \psi ^{\alpha i}  \xi _{\alpha i}\,\,,\,\,\,\,\,\,\,\alpha  = 1,2,3,4,~~i = 
  1,2,...,32,
\lbl{4.21}
\ee
where $\xi_{\tl A}$, $\tl A =1,2,3,...,128$ are the basis spinors\footnote{We now use for the basis spinors and the spinor components in $M_{7,7}$
the same symbols as we did in Eq.\,(\ref{4.16}) for those in $M_{8,8}$.}
in $M_{7,7}$ and $\psi^{\tl A}$ the spinor components. If the manifold $M_{7,7}$ is flat, then
$\xi_{\tl A}$ can be written as the product $\xi_{\tl A} = \xi_\alpha \eta_i$  of the
basis spinors $\xi_\alpha$ in $M_{1,3}$ and the basis spinors $\eta_i$ in the internal space $M_{6,4}$.
 
The spinor $\Psi$ depends on position in $M_{8,8}$ and hence also on position in the subspace $M_{7,7}$.
From now on, $x^M$ will denote coordinates in $M_{7,7}$, and the index ${M}$ will take values $1,2,3,...,14$.
By splitting the coordinates into those of spacetime $M_{1,3}$, and those of the
internal space, we have
\be
\Psi  = \Psi (X^M )\,,\,\,\,\,\,\,\,X^M  = (x^\mu  ,x^{\bar M} )\,,\,\,\,\,\,\mu  = 0,1,2,3;\,\,\,\,\bar M = 5,6,...,14
\lbl{4.22}
\ee
Considering the Dirac equation we find
\be
  (i \gamma ^M \partial _M  + \kappa )\Psi  = i(\gamma ^\mu  \partial _\mu   + \gamma ^{\bar M} \partial _{\bar M}  + \kappa )\Psi   = 0
\lbl{4.23}
\ee
 If the higher dimensional mass $\kappa$ is zero, then the Dirac equation becomes
\be
  \gamma ^M \partial _M \Psi  = \gamma ^\mu  \partial _\mu  \Psi  + \gamma ^{\bar M} \partial _{\bar M} \Psi = 0
\lbl{4.24}
\ee
For the higher dimensional spinor field $\Psi$ we can take the ansatz
\be
\Psi (x^\mu  ,x^{\bar M} ) = \Psi ^{(4)} (x^\mu  )\exp \,[ - ip_{\bar M} x^{\bar M} ]\,{\rm{U}}^{(10)} ,
\lbl{4.25}
\ee
which is the product of a spinor field in $M_{1,3}$ and a spinor field in $M_{0,10}$. This is a simplified model in which the spinor field in the ten dimensional internal space is just a plane wave spinor.
With such ansatz the Dirac equation (\ref{4.24}) becomes
\be
\gamma ^\mu  \partial _\mu  \Psi ^{(4)} U^{(10)}  + \Psi ^{(4)} ( - i)\gamma ^{\bar M} p_{\bar M} {\rm{U}}^{(10)}  = 0
\lbl{4.26}
\ee
Using the relation
\be
\gamma ^{\bar M} p_{\bar M} {\rm{U}}^{(10)}  = m\,{\rm{U}}^{(10)} 
\lbl{4.27}
\ee
we obtain
\be
i\gamma ^\mu  \partial _\mu  \Psi ^{(4)} {\rm{U}}^{(10)} \, + \,m\,\Psi ^{(4)} {\rm{U}}^{(10)}  = 0
\lbl{4.28}
\ee
which is the massive Dirac equation in four dimensions.

A more general ansatz is
\be
\Psi (x^\mu  ,x^{\bar M} ) = \Psi ^{(4)} (x^\mu  )\Psi ^{(10)} (x^{\bar M} )
\lbl{4.29}
\ee
which is not constrained to plane waves. Then from Eqs.\,(\ref{4.24}) and (\ref{4.29})
we obtain
\be
  i\gamma ^{\bar M} \partial _{\bar M} \Psi ^{(10)}  = m\,\Psi ^{(10)},
\lbl{4.30}
\ee
and
\be
i\gamma ^\mu  \partial _\mu  \Psi ^{(4)} \Psi ^{(10)} \, + \,m\,\Psi ^{(4)} \Psi ^{(10)}  = 0 .
\lbl{4.31}
\ee
According to Eq.\,(\ref{4.30}), the internal states are eigenstates of the operator
$ i\gamma ^{\bar M} \partial _{\bar M}$ with $m$ being an eigenvalue. The mass term in the 4D Dirac equation (\ref{4.31}) thus comes from the internal space. This is a well know feature of
Kaluza-Klein theories which work in curved higher dimensional manifolds. Therefore, in the following we will assume 
that $M_{7,7}$ is curved and such that $M_{7,7}= M_{1,3} \dot + M_{6,4}$ still holds locally.

The operator $i\gamma ^{\bar M} \partial _{\bar M}$ is responsible for generations.
When acting on the spinor field $\Psi ^{(10)}  = \psi ^i \eta _i$
it consists of two parts:
\be
\gamma ^{\bar M} \partial _{\bar M} \Psi ^{(10)}  = \left( {\gamma ^{\bar M} \partial _{\bar M} \psi ^i  + \gamma ^{\bar M} \Gamma _{\bar M\,j}^{\,\,i} \psi ^j } \right)\eta _i \equiv \DD_{\bar M} \psi^i \eta_i.
\lbl{4.32}
\ee
The first part is the ``orbital'' contribution, while the second part comes from the action of
the derivative\footnote{
   In ref.\ci{PavsicKaluzaLong} it was explained that the same symbol $\p_M$ can be used for the derivative operator acting on different kinds of objects. For instance, if acting on a scalar field, then it behaves as partial derivative, if acting on a basis vector it gives the connection, and if acting on a basis spinor it give the spin connection. Usage of the same symbol for the derivative in all such cases much simplifies the calculations.}
$\p_{\bar M}$ on the basis spinors $\eta_i$ of the internal space
\be
\partial _{\bar M} \eta _i  = \Gamma _{\bar M\,i}^{\,\,j} \eta _j  .
\lbl{4.33}
\ee
We see that the second part in Eq.\,(\ref{4.32}) is the contribution due to Yukawa coupling, where 
the spin connection in the “internal” space, $\Gamma _{\bar M\,i}^{\,\,j}$,
has the role of {\it Higgs fields}. 

 The eigenvalue equation (\ref{4.30}) in the internal space yields the
possible values of the particle mass $m$. According to Eq.\,(\ref{4.32}), mass does not arise from
the Higgs field only, but also from the orbital momentum $\gam^{\bar M} \p_{\bar M} \psi^i$. Namely, if we multiply
Eq.\,(\ref{4.32}) from the left by $-i \gam^{{\bar N}} \p_{\bar N}$ and then also by ${\eta^{j}}^\dg$,
we obtain, after taking the scalar part and renaming the indices, the following
equation\ci{PavsicBook2}:
\be
  g^{\bar M \bar N} \DD_{\bar M}\DD_{\bar N} \psi^i +{(\sg_{\bar M \bar N})^i}_j {R_{\bar M \bar N}^j}_k \psi^k
  = m^2 \psi^i ,
\lbl{4.33}
\ee
where
\be
  g^{\bar M \bar N} \DD_{\bar M}\DD_{\bar N} \psi^i = 
   g^{\bar M \bar N}\left (\p_{\bar M} \DD_{\bar N} \psi^i
  - \Gam_{\bar M \bar N}^{\bar K} \DD_{\bar K} \psi^i + \Gam_{\bar M j}^i \p_{\bar N} \psi^j \right ).
\lbl{4.34}
\ee
and
\be
  g^{\bar M \bar N} \p_{\bar M} \DD_{\bar N} \psi^i 
  = g^{\bar M \bar N} \p_{\bar M}\left ( \p_{\bar N} \psi^i  + {\Gam_{\bar N}^i}_j \psi^j \right ).
\lbl{4.35}
\ee
Here ${(\sg_{\bar M \bar N})^i}_j = \frac{1}{2}{([\gam_{\bar M},\gam_{\bar N}]^i}_j$
is the spin tensor
and ${R_{\bar M \bar N}^i}_k$ the curvature tensor expressed in term of the spin connection.
The term $g^{\bar M \bar N} \p_{\bar M}\p_{\bar N} \psi^i$, written in terms of the spherical coordinates
in the internal space, becomes an expression that contains the ``orbital'' momentum operator
acting on the internal state $\psi^i$. In this setup the number of generations is given by the number of eigenstates and eigenvalues determined by the equation (\ref{4.30}). Moreover, the fact that masses are not determined
by Higgs fields only, but also by the orbital momentum in the internal space should be taken into account and recalculate, e.g., the proton life time.

In general, in Eq.\,(\ref{4.24}), the derivative $\p_M$, acting on the basis spinors $\xi_{\tl A}$, gives the
connection $\Gam_{M \tl B}^{~\tl A}$ that includes the ordinary spin connection, a Yang-Mills field and a multiplet of Higgs fields. All those fields are contained in the covariant derivative $\p_M \Psi = (\p_M \psi^{\tl A} +
\Gam_{M \tl B}^{~\tl A} \psi^{\tl B})\xi_{\tl A}$. A first step in this direction was proposed in Ref.\ci{PavsicKaluzaLong},
where the internal piece of the connection was not yet recognized as a Higgs multiplet. That the covariant Dirac
derivative contains all those fields, including the Higgs multiplet, was observed in Ref.\ci{Lisi2}, but without using the concept of geometric
spinors defined according to Eq.\,(\ref{4.21}).

\section{Conclusion}

Fundamental objects such as strings and branes can be described in terms of their spacetime
volume degrees of freedom represented by Clifford numbers belonging to $Cl(1,3)$.
The latter algebra is considered as a tangent space of a 16-dimensional manifold,
called Clifford space.
Clifford algebra as a vector space can be considered as being
spanned over 16 basis vectors that are generatore of $Cl(8,8)$
or its complexified version.
We thus have a 16-dimensional vector space $V_{8,8}$ that have enough
room for the unification of fundamental particles and interactions
including gravity. Spinors are members of left ideals of $Cl(8,8)$.
They satisfy the Dirac equation in sixteen dimensions.
The extra dimensions give rise not only to the presence of the first
generation particles of the $SO(10)$ grand unification, but also
to  different generations. Namely, the effective mass in four dimensions
is due to the presence of extra dimensions, a known feature of Kaluza-Klein
theories. But in those theories there is also a problem of how to reconcile the chiral properties
of spinors in higher dimensions with the observed parity non conservation in four dimensions.
Within the framework considered here, such problem does not exist. Namely, considering the concept of Clifford space whose tangent space at any of its points is a Clifford algebra, with its members being spinors, automatically leads\ci{PavsicSpinorInverse,PavsicMoskva,PavsicLicata} to
mirror particles\ci{LeeYang,Kobzarev,PavsicMirror,Kolb,Blinnikov1,Blinnikov2,Foot1}
that are coupled to mirror gauge fields and are unobservable by means of ordinary gauge fields. Therefore, as extensively investigated in Refs.\ci{Foot1,Foot2,Foot3,Foot4,Hodges,Foot5,Foot6,Mohapatra,Berezhiani,Ciarcelluti1,Ciarcelluti2,Ciarcelluti3,FootReview,PavsicMoskva,PavsicLicata},
they are candidates for dark matter.


\begin{thebibliography}{12}

\bi{PavsicBook1} M.~Pav\v si\v c,
{\it The Landscape of theoretical physics: A Global view. From point particles to the brane world and beyond, in search of a unifying principle}, (Kluwer Academic, 2001)
[arXiv:gr-qc/0610061 [gr-qc]].

\bi{PavsicBled} M.~Pav\v si\v c,
``Towards a new paradigm: Relativity in configuration space,''
Contribution to: Time and Matter 2007, 161-178
[arXiv:0712.3660 [gr-qc]].

\bi{PavsicIARD2016} M.~Pav\v{s}i\v{c},
``Branes and Quantized Fields,''
{\it J. Phys. Conf. Ser.} \textbf{845}, no.1, 012018 (2017)
doi:10.1088/1742-6596/845/1/012018
[arXiv:1703.05140 [hep-th]].

\bi{PavsicBook2} M.~Pav\v{s}i\v{c},
{\it Stumbling Blocks Against Unification: On Some Persistent Misconceptions in Physics},
(World Scientific, 2020),
doi:10.1142/11738.

\bi{CastroChaos} C. Castro, {\it Chaos, Solitons and Fractals} 
{\bf 10}, 295 (1999);
{\bf 11}, 1663 (2000);{\bf 12}, 1585 (2001);\\
C. Castro,  {\it Found. Phys.} {\bf 30}, 1301 (2000).

\bi{CastroAurilia} S. Ansoldi, A. Aurilia, C. Castro and E. Spallucci, 
{\it Phys. Rev. D} {\bf 64}, 026003 (2001) [hep-th/0105027].

\bi{Aurilia} A. Aurilia, S. Ansoldi and E. Spallucci, {\it 
	Class. Quant. Grav.} {\bf 19},  3207 (2002).

\bi{PavsicArena} M.~Pav\v si\v c,
{\it Found. Phys.} \textbf{33}, 1277-1306 (2003)
doi:10.1023/A:1025637126758
[arXiv:gr-qc/0211085 [gr-qc]].

\bi{PavsicKaluza} 
M.~Pav\v si\v c,
{\it Phys. Lett. B} \textbf{614}, 85-95 (2005)
doi:10.1016/j.physletb.2005.03.052
[arXiv:hep-th/0412255 [hep-th]].

\bi{PavsicKaluzaLong} M. Pav\v si\v c,
{\it Int.\ J.\ Mod.\ Phys.}\ A {\bf 21}, 5905 (2006)
[arXiv:gr-qc/0507053].


\bi{CastroPavsicReview} C. Castro and M. Pav\v si\v c,
\emph{Prog.\ Phys.} {\bf 1}, 31 (2005).

\bi{PavsicLicata} M. Pav\v si\v c, "Quantized Fields \`a la Clifford and Unification",
in {\it Beyond Peaceful Coexistence;
	The Emergence of Space, Time and Quantum} (Ignazio Licata, Ed.,
Imperial College Press, 2016), pp. 615--660,
[arXiv:1707.05695 [physics.gen-ph]].

\bibitem{PavsicMaxwellBrane}
M.~Pav\v si\v c,
{\it Found. Phys.} \textbf{37}, 1197-1242 (2007)
doi:10.1007/s10701-007-9147-3
[arXiv:hep-th/0605126 [hep-th]].

\bi{PavsicE8} M.~Pav\v si\v c,
{\it J. Phys. A} \textbf{41}, 332001 (2008)
doi:10.1088/1751-8113/41/33/332001
[arXiv:0806.4365 [hep-th]].

\bi{Pati-Salam} J. C. Pati, A. Salam, {\it Phys. Rev. D} {\bf 10} (1), 275 (1974),
doi:10.1103/physrevd.10.275.

\bi{BaezPatiSalam} J. C. Baez, J. Huerta (2009). "The Algebra of Grand Unified 
Theories". arXiv:0904.1556 [hep-th].


\bibitem{Stueckelberg1}   E. C. G. Stueckelberg, {\it Helv. Phys. Acta} {\bf 14}, 322 (1941).  

\bibitem{Stueckelberg2}  E.C.G. Stueckelberg, {\it Helv. Phys. Acta} {\bf 15}, 23 (1942). 

\bibitem{Horwitz1}  L.P. Horwitz and C. Piron, {\it Helv. Phys. Acta} {\bf 46}, 316 (1973).

\bibitem{Fanchi} J.R. Fanchi,  {\it Found. Phys.} {\bf 23}, 287 (1993), and many references therein.

\bi{Bars}
I. Bars, C. Deliduman, and O. Andreev, {\it Phys. Rev. D} {\bf 58}, 066004 (1998) ;\\
I. Bars, {\it Phys. Rev. D} 58 (1998) 066006;\\
I. Bars, {\it Class. Quant. Grav.} {\bf 18} (2001) 3113; I. Bars, {\it Phys. Rev. D} {\bf 74}, 085019 (2006).

\bi{Lisi1} A.~G.~Lisi,
``An Explicit Embedding of Gravity and the Standard Model in E8,''
[arXiv:1006.4908 [gr-qc]].

\bi{Lisi2} A.~G.~Lisi, L.~Smolin and S.~Speziale,
``Unification of gravity, gauge fields, and Higgs bosons,''
J. Phys. A \textbf{43}, 445401 (2010)
doi:10.1088/1751-8113/43/44/445401
[arXiv:1004.4866 [gr-qc]].

\bi{Gillard}  A.~B.~Gillard and N.~G.~Gresnigt,
[arXiv:1906.05102 [physics.gen-ph]].

\bi{Gresnigt}
N.~G.~Gresnigt,
``The Standard Model particle content with complete gauge symmetries from the minimal ideals of two Clifford algebras,''
Eur. Phys. J. C \textbf{80}, no.6, 583 (2020)
doi:10.1140/epjc/s10052-020-8141-1
[arXiv:2003.08814 [physics.gen-ph]].

\bi{CastroUnif1} C.~Castro,
``Clifford Algebraic Unification of Conformal Gravity with an Extended Standard Model,''
Adv. Appl. Clifford Algebras \textbf{27}, no.2, 1031-1042 

(2017)
doi:10.1007/s00006-016-0702-x

\bi{CastroUnif2} C.~Castro Perelman,
``R $\otimes $ C $\otimes $ H $\otimes $ O-Valued Gravity as a Grand Unified Field Theory,''
Adv. Appl. Clifford Algebras \textbf{29}, no.1, 22 (2019)
doi:10.1007/s00006-019-0937-4

\bi{CastroUnif3}C.~Castro Perelman,

Hermitian Matrix Geometry and Nonsymmetric Kaluza\textendash

{}Klein Theory,''
Adv. Appl. Clifford Algebras \textbf{29}, no.3, 58 (2019)
doi:10.1007/s00006-019-0977-9

\bi{CastroUnif4} C. Castro On Jordan-Clifford Algebras, Three
Fermion Generations with Higgs Fields
and a SU(3) x SU(2)L x SU(2)R x U(1) (Available on Academia.edu)

\bibitem{PavsicLocalTachyons}
M.~Pav\v si\v c,
{\it Adv. Appl. Clifford Algebras} \textbf{23}, 469-495 (2013)
doi:10.1007/s00006-013-0381-9
[arXiv:1201.5755 [hep-th]].

\bi{PavsicBg2008} M. Pav\v si\v c,
"Beyond spacetime: On the Clifford algebra based generalization of relativity", 
Contribution to 5th Summer School in Modern Mathematical Physics (MPHYS5)
6-17 July 2008.
Belgrade, Serbia, https://inspirehep.net/conferences/979332,
(C08-07-06.1)     
pp. 343-356  [https://arxiv.org/pdf/0907.2773].

\bi{CastroUnific} C. Castro, "A Clifford Algebra Based Grand Unification Program of Gravity and the Standard Model : A Review Study",
{\it Can.J.Phys.} {\bf 92},1501-1527 (2014) 
DOI: 10.1139/cjp-2013-0686 .

\bi{PavsicSaasFee}
M.~Pav\v si\v c,
{\it Found. Phys.} {\bf 35}, 1617 (2005)
doi:10.1007/s10701-005-6485-x
[arXiv:hep-th/0501222 [hep-th]].

\bi{Cangemi} D. Cangemi, R. Jackiw and B. Zwiebach, {\it Annals of Physics} {\bf 245}, 408 (1996).

\bi{Jackiw} E. Benedict, R. Jackiw and H. J. Lee, {\it Phys. Rev.} D. {\bf 54}, 6213 (1996).

\bi{Kim} Y.S. Kim and M.E. Noz, {\it Phys. Rev. D} {\bf 8}, 3521 (1973); {\bf 12}, 122 (1975); 

{\bf 15},335 (1977); {\it Phys. Rev. Lett.} {\bf 63}, 348 (1989).

\bi{KimBook} Y.S. Kim and M.E. Noz, Theory
and Applications of the Poincare\'e Group (D. Reidel Publishing Company, Dordrecht, 1986).

\bi{PavsicPseudoHarm} M. Pav\v si\v c,
{\it Phys.\ Lett.}\ A {\bf 254}, 119 (1999)
[arXiv:hep-th/9812123].

\bibitem{Woodard} 
R.~P.~Woodard,
{\it Lect.\ Notes Phys.}\  {\bf 720}, 403 (2007)
[arXiv:astro-ph/0601672].

\bi{Coleman-Mandula} S.  Coleman, J. Mandula, {\it Phys. Rev.} {\bf 159} (5), 1251 (1967) 
Bibcode:1967PhRv..159.1251C. doi:10.1103/PhysRev.159.1251.


\bi{SpinorFock} E. Cartan, {\it Le\c cons sur la th\' eorie des
	spineurs I \& II} (Paris: Hermann, 1938)\\
E. Cartan  {\it The theory of spinors}, English transl. by R.F Streater,
(Paris: Hermann, 1966)\\
C. Chevalley {\it The algebraic theory of spinors}
(New York: Columbia U.P, 1954 )\\
I .M. Benn, R .W. Tucker, {\it An introduction to spinors and geometry with
	appliccations in physics} (Bristol: Hilger, 1987).

\bi{BudinichP} 
P. Budinich  {\it Phys. Rep.} {\bf 137}, 35 (1986);\\
P. Budinich and  A. Trautman  {\it Lett. Math. Phys.} {\bf 11}, 315 (1986).

\bi{Giler} S. Giler,  P. Kosi\' nski, J. Rembieli\' nski and P. Ma\' slanka, 
{\it Acta Phys. Pol. B}  {\bf 18}, 713 (1987).

\bi{Winnberg} J. O. Winnberg,  {\it J. Math. Phys.} {\bf 18},  625 (1977).

\bi{PavsicSpinorInverse} M. Pav\v si\v c,
Behavior in Weak Interactions,''
{\it Phys.\ Lett.} B {\bf 692}, 212 (2010)
[arXiv:1005.1500 [hep-th]].

\bibitem{PavsicOrthoSymp}
M.~Pav\v si\v c,
{\it v. Appl. Clifford Algebras} \textbf{22}, 449-481 (2012)
doi:10.1007/s00006-011-0314-4
[arXiv:1104.2266 [math-ph]].

\bi{PavsicFirence}
M.~Pav\v si\v c,
{\it J. Phys. Conf. Ser.} {\bf 437}, 012006 (2013)
doi:10.1088/1742-6596/437/1/012006
[arXiv:1210.6820 [hep-th]].

\bi{BudinichM} M. Budinich, {\it J. Math. Phys.}, {\bf 50}, 053514 (2009).\\
M. Budinich,  {\it J. Phys. A}, {\bf 47}, 115201 (2014).\\
M. Budinich, {\it Adv. Appl. Clifford Algebras}, {\bf 25}, 771 (2015).

\bibitem{PavsicMoskva}
M.~Pav\v{s}i\v{c},
``Geometric Spinors, Generalized Dirac Equation and Mirror Particles,''
Contribution to 3rd International Conference on Theoretical Physics (ICTP 2013)
24-28 June 2013. Moscow, Russian Federation
(C13-06-24.8)
[arXiv:1310.6566 [hep-th]].

\bi{LeeYang}  T. D. Lee  and C. N. Yang,  {\it Phys. Rev.}, {\bf 104}, 254 (1956).

\bi{Kobzarev} I. Yu. Kobzarev, L. B. Okun and I. Ya. Pomeranchuk,
{\it Soviet J. Nucl. Phys.} {\bf 5},  837 (1966).

\bi{PavsicMirror} M. Pav\v si\v c, {\it Int. J. Theor. Phys.} {\bf 9},
229 (1974).

\bi{Kolb} E.W. Kolb, D. Seckel, M.S. Turner {\it Nature} {\bf 314}, 415 (1985).

\bi{Blinnikov1}  S. I. Blinnikov and  M. Y. Khlopov, {\it Sov. J. Nucl. Phys.}
{\bf 36}, 472 (1982) [{\it Yad. Fiz.}, {\bf 36}, 809 (1982)].

\bi{Blinnikov2} S. I. and M. Y. Khlopov, {\it Sov.Astron.}, {\bf 27}, 371  (1983) 
[{\it Astron. Zh.}, {\bf 60}, 632 (1983)].


\bi{Foot1} R. Foot, H. Lew and R. R. Volkas {\it Phys. Lett.} B {\bf 272}, 67 (1991).

\bi{Foot2} R. Foot, H. Lew and R. R. Volkas {\it Mod. Phys. Lett.} A {\bf 7}, 2567 (1992).

\bi{Foot3} R. Foot {\it Mod. Phys. Lett.} {\bf 9}, 169 (1994).

\bi{Foot4} R. Foot and R. R. Volkas, {\it Phys. Rev. D} {\bf 52}, 6595 (1995).

\bi{Hodges} H. M. Hodges, {\it Phys. Rev. D} {\bf 47}, 456 (1993).

\bi{Foot5} R. Foot, {\it Phys. Lett. B} {\bf 452}, 83 (1999).

\bi{Foot6} R. Foot, {\it Phys. Lett. B} {\bf 471}, 191 (1999).

\bi{Mohapatra} R.N. Mohapatra, {\it Phys. Rev. D} {\bf 62}, 063506 (2000).

\bi{Berezhiani} Z. Berezhiani, D. Comelli and F. Villante, {\it Phys. Lett. B}
{\bf 503}, 362 (2001).

\bi{Ciarcelluti1} P. Ciarcelluti, {\it Int. J. Mod. Phys. D} {\bf 14}, 187 (2005).

\bi{Ciarcelluti2} P. Ciarcelluti, {\it Int. J. Mod. Phys. D} {\bf 14}, 223 (2005).

\bi{Ciarcelluti3} P. Ciarcelluti and R. Foot, {\it Phys. Lett. B} {\bf 679}, 278 (2009).

\bi{FootReview}  R. Foot, 
{\it Int.\ J.\ Mod.\ Phys. A}, {\bf 29}, 1430013 (2014)
[arXiv:1401.3965 [astro-ph.O]].



\end{thebibliography}
\end{document}